\documentstyle[aaspp4,11pt,epsf]{article}
\def\bld#1{\mbox{\boldmath$#1$\unboldmath}}

\def\unsetyr{\def\oyear{\relax}\def\cyear{\relax}\def\cyeara{a\relax}\def\cyearb{b\relax}\def\cyearc{c\relax}\def\cyeard{d\relax}}
\def\setyr{\def\oyear{(}\def\cyear{)}\def\cyeara{a)}\def\cyearb{b)}\def\cyearc{c)}\def\cyeard{d)}}
\unsetyr
\def\jcite#1{\setyr\cite{#1}\unsetyr}

\def\rmmat#1{{\hbox{\rm #1}}}
\def\rmscr#1{\rmmat{\scriptsize #1}}
\newcommand{\be}{\begin{equation}}
\newcommand{\ee}{\end{equation}}
\newcommand{\bt}{\begin{table} \begin{center}}
\newcommand{\et}{\end{center} \end{table}}
\newcommand{\ba}{\begin{eqnarray}}
\newcommand{\ea}{\end{eqnarray}}
\newcommand{\ie}{{\it i.e.~}}
\newcommand{\eg}{{\it e.g.~}}

\def\d{{\rm d}}

\def\dd#1#2{\frac{\d #1}{\d #2}}

\newcommand{\comment}[1]{\relax}
\def\eqref#1{Equation~\ref{eq:#1}}

\def\figref#1{Figure~\ref{fig:#1}}

\begin{document}

\title{Probing the Properties of Neutron Stars with Type-I X-Ray
Bursts}

\author{Jeremy S. Heyl}
\authoremail{jsheyl@tapir.caltech.edu}
\affil{
Theoretical Astrophysics,
mail code 130-33,
California Institute of Technology,
Pasadena CA 91125}

\begin{abstract}
The increase in spin frequency as the burning atmospheres of Type~I
X-ray bursts cool provides a strong constraint on the radius of the
underlying neutron star.  If the change in spin frequency is due to a
change in the thickness of the atmosphere, the radius of the star must
exceed $3 G M/c^2$ for any equation of state and approximately $3.5 G
M/c^2$ for most physically reasonable equations of state.  This
constraint arises because the direction of the Coriolis force for
radial motion reverses for $R <  G / c^2 ( M +  I / R^2 )$.
Furthermore, the marked change in the magnitude of the Coriolis force
near compact stars provides a straightforward explanation for why the
frequency of the quickly rotating bursters shifts by the same amount
as the slow rotators; they are slightly more massive, 1.6~M$_\odot$
versus 1.4~M$_\odot$.
\end{abstract}

\section{Introduction}

If material accretes onto the surface of a neutron star sufficiently
slowly, a layer of fuel develops and then suddenly ignites producing a
burst of x-rays known as a Type~I X-ray burst.  These thermonuclear
flashes each release about $10^{39}$~ergs and repeat on a timescale
from hours to days (\cite{Lewi95,Bild98}).  Although only one source
(SAX~J1808.4-3658) exhibits periodic variation in its quiescent
emission (\cite{Wijn98,Chak98}), during the bursts themselves the
emission is quasiperiodic (\eg \cite{Stro97a}), apparently due to
rotational modulation of the inhomogeneities in the thermonuclear
burning. \jcite{1999ApJ...516L..81S} find that the oscillations in the
cooling tails of the X-ray bursts from 4U~1702-429 and 4U~1728-34 are
nearly coherent with $Q\sim 4000$ and consistent with an increase in
the modulation frequency as the burning layer cools.

\jcite{Joss78} calculated the first numerical models of thermonuclear
burning on the surface of a 1.4~M$_\odot$ neutron star with a radius
of 6.6~km.  He found that the initial surface layer of helium expands
from a thickness of about three meters to one of thirty meters during
the onset of the burst (also \cite{Bild95}).  \jcite{Stro97b} argue
that this increase in radius is sufficient to account for the frequency shifts
observed during the bursts due to the conservation of angular momentum.

To use the conservation of angular momentum to explain the change in
frequency of the emission as the burning region of the atmosphere (or
hotspot) expands and then contracts, one assumes that the specific
angular momentum of a fluid element is conserved during the burst.
The magnetic fields of the neutron stars are likely to be weak
(otherwise there would be periodicities in the persistent emission);
consequently, this is a viable assumption.  However, the relationship
between the specific angular momentum of a fluid element and its
position is more complicated in general relativity than in Newtonian
theory (\eg \cite{AbramowiczMillerStuchlik1993}).  In the case of the
Schwarzschild spacetime, the angular momentum of a fluid element is
\be
\d\! L = \d m \Omega r_g^2 = \d m \Omega \frac{r^2}{1-2 M / r}
\label{eq:rgyration}
\ee
where $\Omega$ is the angular velocity of the element as measured by
an observer at infinity.  This expression and its generalizations
result in the reversal of the centrifugal forces (\eg
\cite{AbramowiczPrasanna1990,Abramowicz1993}) and gyroscopic
precession (\eg \cite{1998GReGr..30..593N}) in curved spacetimes.

Rather than using \eqref{rgyration} to obtain the change in the angular
velocity of the hotspot, \S~\ref{sec:coriolis} calculates the change
the angular velocity in a frame that rotates with the neutron star.
The reasons for this treatment are two-fold.  First, the calculation
is quite straightforward and illustrative.  Second, the effects of
frame dragging are likely to be important for neutron stars, so 
rather than deriving a new definition for the radius of gyration to
include frame dragging, frame dragging may simply be included in
the metric.  \S~\ref{sec:xrayburst} applies these results to Type~I
X-ray bursts, and finally \S~\ref{sec:conclus} summarizes the conclusions.

\section{The Coriolis Force in Curved Spacetime}
\label{sec:coriolis}

Evaluating the Coriolis force in the curved spacetime surrounding a
neutron star is straightforward once the metric is specified.  The
forces that affect the motion of
the neutron star atmosphere are most easily expressed in a frame that
rotates with the star.   In this frame the nonzero components of the metric
tensor are
\ba
g_{00} &=&  1 - \frac{2M}{r} + \Omega^2 r^2 \sin^2\theta \left
( \frac{I}{r^3} + 1 \right ) \\
g_{03} = g_{30} &=& \Omega r^2 \sin^2 \theta \left ( 2 \frac{I}{r^3} + 1 \right
) \\
g_{11} &=& - \left ( 1 - \frac{2M}{r} \right )^{-1} \\
g_{22} &=& - r^2 \\
g_{33} &=& -r^2 \sin^2 \theta.
\ea
where $x^\mu=\left [ t, r, \theta, \phi \right ]$.
$J=I\Omega$ which accounts for the effects of frame dragging to
lowest order in $\Omega$ (\cite{Land2}), and the various quantities
are given in geometric units with $c=G=1$.  For rotational frequencies
, $\nu<400$~Hz, this is a good approximation
(\eg \cite{1999ApJ...513..827M}) but for $\nu \sim
600$~Hz, the induced quadrupole moment of the star may introduce a
significant but not dominant correction to the metric
(\cite{1994ApJ...424..823C,1998AstL...24..774S}); this issue will be
addressed in the more detailed sequel.

The Coriolis force for material moving radially manifests itself
through the connection term, 
\ba
\Gamma^{3}_{01} &=& \frac{1}{2} \left ( g^{33} g_{30,1} + g^{30} g_{00,1} \right ) \\
                &=&  \frac{\Omega}{r} \left [ \left ( 1 -
\frac{I}{r^3} \right ) +
\frac{1 + 2 I/r^3}{1 - 2 M/r} \frac{M}{r} \right ],
\ea
and $\Gamma^{3}_{11}=\Gamma^{3}_{00}=0$.
The first term yields the standard ``Newtonian'' Coriolis force
reduced by the effects of frame dragging, and the second term arises
because the usually afferent gravitational acceleration has an azimuthal
component for material moving radially in the rotating frame.

The atmosphere expands with a four velocity $\bld{v}$ whose components
in the rotating frame are found by setting $\bld{u \cdot u}=1$,
$\bld{v \cdot v}=1$ and $\bld{u \cdot v}=\gamma$ where $\bld{u}$
points along the time axis in the rotating frame (\ie it is four
velocity of an observer stationary in the rotating frame).
Due to pressure gradients, the trajectory of gas does not obey the
$t$ and  $r$ components of the geodesic equations.  In the
non-rotating frame, these pressure gradients also have a $\phi$
component.  However, in the rotating frame, the gas is free to
follow a geodesic in the $ \phi$ direction, yielding
\be
\dd{v^3}{\tau} = -2 \beta \gamma^2 \frac{\Omega}{r}
\left ( 1 - \frac{3 M}{r} - \frac{I}{r^3}  \right )
\left ( 1 - \frac{2 M}{r} \right )^{-1} + {\cal O} \left (\Omega^3 \right )
\ee
where $\beta$ is the magnitude of the three-velocity measured locally
by an observer stationary in the rotating frame (\ie ${\bld u}$) and
$\gamma = 1/\sqrt{1-\beta^2}$.  

If the radial motion is nonrelativistic and small relative to the
initial radius (as for the expansion of the atmosphere), the result is
easily integrated.  After transforming the result to the nonrotating
(distant observer's) frame, it is
\be
\dd{\ln \Omega_\infty}{\ln r} = -2 \left ( 1 - \frac{3 M}{r} - \frac{I}{r^3}  \right )
\left ( 1 - \frac{2 M}{r} \right )^{-1}.
\label{eq:dlnOmega}
\ee
This result in the absence of frame dragging (\ie $I\rightarrow 0$) could
also be obtained by using the definition of
\jcite{AbramowiczMillerStuchlik1993} for the radius of gyration in
general relativity.  Comparison of this result with \eqref{rgyration}
yields the following definition for the radius of gyration with frame
dragging included to lowest order in $\Omega$,
\be
r_g = \frac{r}{\sqrt{1-2M/r}} \left ( 1 - \frac{2 M}{r} \right
)^{-\frac{1}{8} \frac{I}{M^3} } \exp \left ( -\frac{1}{4} \frac{r +
M}{M^2 r^2} I \right )
\ee
As the mass of the star and its moment of inertia vanish, the familiar
result obtains.  However, in the vicinity of a neutron star, the
correction may be large.

\section{Application to X-Ray Bursts}
\label{sec:xrayburst}

\jcite{1999ApJ...516L..81S} find that during a X-ray burst, the
observed frequency of the modulation exponentially decreases toward
a value which is constant from burst to burst for each source.  They
interpret this asymptotic frequency as the spin frequency of the star.
At the beginning of the burst, they argue that the atmosphere expands
and slows due to the Coriolis force, and as it cools and deflates the
spin frequency increases again through the Coriolis force.  This
simple inverse correlation between the spin frequency and the
thickness of the atmosphere restricts
\be
R > 3 M + \frac{I}{R^2}
\ee
for the neutron stars associated with 4U~1636-54, KS~1731-26, Aql X-1,
the galactic center source, 4U~1702-43 and 4U~1728-34 ( $R$ and $M$
are the mass and radius of the neutron star).  For a uniform density
sphere in the Newtonian limit $I=\frac{2}{5} M R^2$.
\jcite{1994ApJ...424..846R} find that the moment of inertia for the
FPS equation of state (\cite{Frie81}) is well fit by
\be
I = 0.21 M R^2 \left ( 1 - \frac{2 M}{R} \right )^{-1}.
\ee
In the context of this equation of state, $R>3.49 M$.

\jcite{Cumm00} have examined the structure of the atmosphere of the
X-ray burst in further detail.  By understanding the magnitude of the
change in the frequency of X-ray bursts as well as its sign, tighter
constraints on the radius of the neutron star may be obtained.  The
the proper thickness of the layer ($\Delta z_{1.89}$) as calculated by
\jcite{Cumm00} for $g_s=1.89\times 10^{14}$~cms$^{-2}$ and the coordinate
thickness ($\Delta r$) are related by
\be
\frac{\Delta r}{R} = 2.1 \times 10^{-7} \rmmat{cm}^{-1} \Delta z_{1.89}
\frac{R}{M} \left( 1 - \frac{2 M}{R} \right )
\label{eq:deltar}
\ee
where the variation in the surface gravity with mass and radius is
included and $2.1 \times 10^{-7} \rmmat{cm}^{-1} = g_s c^2$.
The change in observed frequency $\Delta \nu$ for $\nu=300$~Hz is
\be
\Delta \nu = -0.252~\rmmat{Hz} \frac{\Delta z_{1.89}}{20~\rmmat{m}}
\frac{\nu}{300~\rmmat{Hz}} \frac{R}{M}
\left ( 1 - \frac{3 M}{R} - \frac{I}{R^3}  \right )
\ee

It is also straightforward to add the relativistic corrections to the
more detailed results of \jcite{Cumm00} (C\&B) which yields
\be
\left . \frac{\Delta \Omega_\infty}{\Omega_\infty} \right |_\rmscr{GR}
=  \frac{R}{M}
\left ( \left . \frac{R}{M} \right |_\rmscr{C\&B}
 \right)^{-1}
\left ( 1 - \frac{3 M}{R} - \frac{I}{R^3}  \right )
\left . \frac{\Delta \Omega}{\Omega} \right |_\rmscr{C\&B}
\label{eq:cummcorrect}
\ee
where the effects of changing surface gravity are also included by
assuming that the scale height of the atmosphere is inversely
proportional to the gravitational acceleration;
\jcite{Cumm00} assume that $R/M\approx 4.82$.

\figref{correct} shows the corrections to the calculated values of
$\Delta \Omega$ including both the effects of general relativity and
varying the surface gravity.  If the effects of the curved spacetime
are included, the result is given by the middle curve, and if frame
dragging is also included the lower curve traces the result.  The 
expected change
in the observed angular velocity as the atmosphere expands vanishes
for $R=3 M$ (without frame dragging) and $R=3.49 M$ (with frame dragging).
\begin{figure}
\plotone{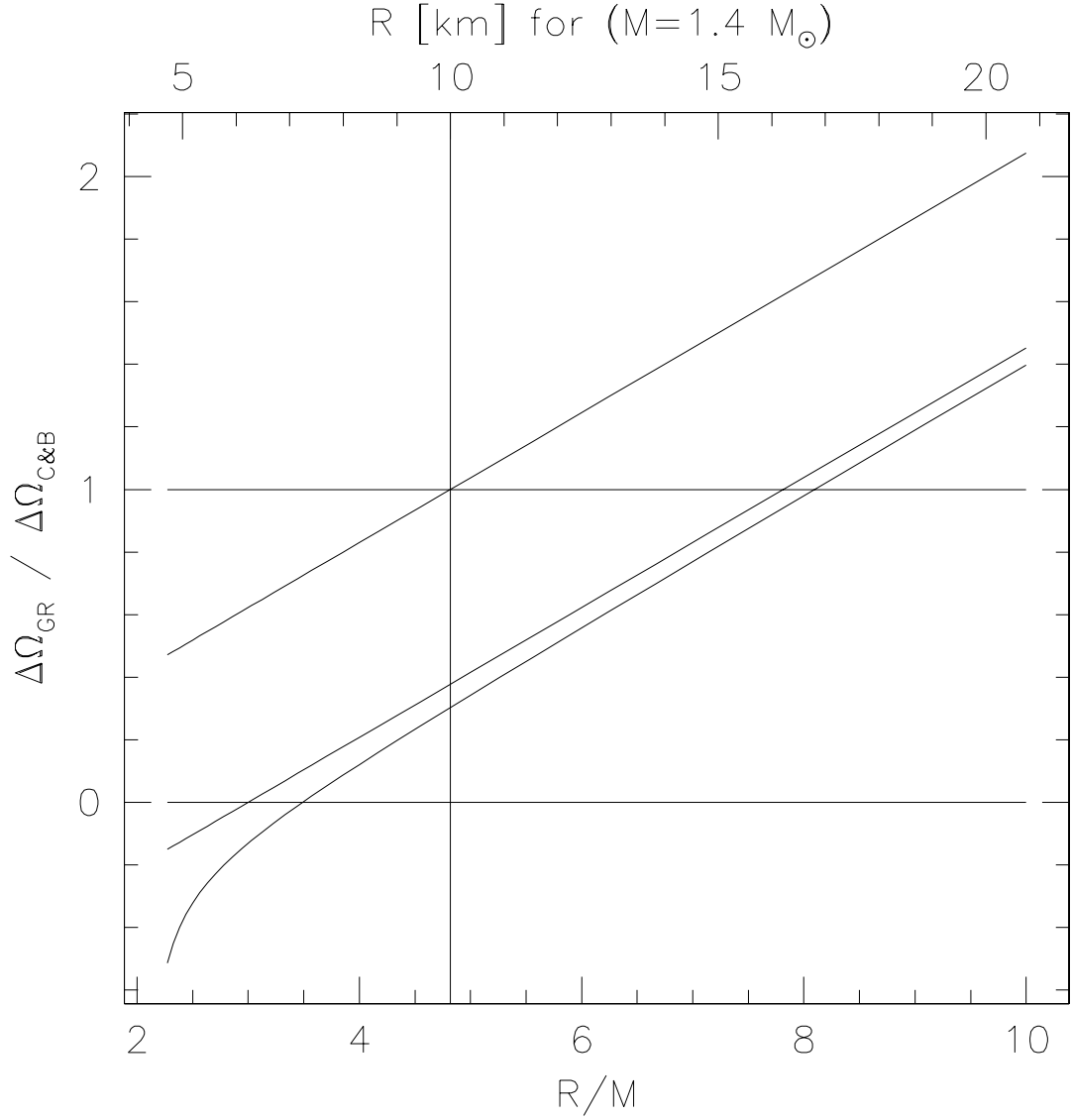}
\caption{The relativistic corrections to the expected spin-down from
Type~I X-ray bursts.  The upper slanted line gives the Newtonian
result, and the cross shows that it is normalized for a neutron star
with $R=10$~km and $M=1.4$~M$_\odot$.  The middle line gives the
relativistic correction without frame dragging, and the lower curve
traces the correction including frame dragging.}
\end{figure}

Table~1 of \jcite{Cumm00} presents the properties of the oscillations
from Type~I bursts.  Specifically, the sources fall into two types
those with $\nu \sim 300$~Hz and those with $\nu \sim 600$~Hz.  Both
sets of objects typically have $\Delta \nu \sim 1$~Hz.
\jcite{Mill99} argues that those objects with faster oscillations
simply have two burning fronts active simultaneously, so the actual
rotational frequency of the star is $\sim 300$~Hz and the kinematics of
the expanding atmosphere is similar to those with faster
oscillations.   An alternative possibly is that the 600-Hz
oscillations correspond to a 600-Hz rotational frequency and that
$\Delta \nu/\nu$ is smaller due to a different value of $M/R$.

To wit, assuming that the 300-Hz objects are 1.4 M$_\odot$ neutron
stars with the FPS equation of state
(\cite{Frie81,1994ApJ...424..846R}) they have a radius of 10.75~km and 
$M/R=0.196$.  Using this data to normalize \eqref{cummcorrect} 
for $\Delta \nu/\nu = 6 \times 10^{-3}$ gives
\be
\left . \frac{\Delta \Omega}{\Omega} \right |_\rmscr{C\&B} = 13.5
\times 10^{-3}
\label{eq:cummnorm}
\ee
which is about twice as large as the value that \jcite{Cumm00} find
for their Newtonian models.

Using this value, \eqref{cummcorrect} can be used to infer a
value of $M/R$ for the fast rotators of 0.234, modestly different 
from that of the slow rotators (0.246 if frame dragging is neglected).
Without the general relativistic corrections, $M/R$ for the fast
rotators would be 0.391 which is more compact than is achieved for any
standard equation of state.  Using the value of $M/R$ which includes
frame dragging yields $M=1.6 \rmmat{M}_\odot$ for the fast rotators
and $R=10.3$~km.

If more detailed models argue that the normalization
(\eqref{cummnorm}) is unphysically large, the conclusion would
be that the objects have $R \gg 3 M$. Specifically, if the modeling of
the evolving atmosphere by \jcite{Cumm00} is precisely correct,
$R = 8.1 M$ for the slowly rotating bursters and $R=5.7 M$ for the fast
rotators.  If the mass of the neutron stars exceeds a solar mass, this
requires a very hard equation of state such as the TI and MF
results (\cite{1976ApJ...208..550P}).

\section{Conclusions}
\label{sec:conclus}

The Coriolis force in the curved spacetime near a neutron star is
substantially smaller than the Newtonian value and reverses its
direction as the radius of the star approaches $3 G M/c^2$.  For
typical neutron-star parameters of $R=10$~km and $M=1.4$~M$_\odot$,
the fully relativistic Coriolis force is only 30\% of the Newtonian
value.  This contrasts with the assertion of \jcite{Stro99} that the
correction to the Coriolis force is the same order as the
gravitational redshift, \ie 20 -- 30\% due to time dilation alone.  
Although time dilation does contribute,  it does not dominate.  Time
dilation alone would not predict, the change in the sign of the
Coriolis force for $r\approx 3 M$.  

The bulk of the effect arises from the fact that the specific angular
momentum of a fluid element has a minimum at the photon circular
orbit, which has $r=3 M$ for a non-rotating star.  This also
manifests itself through the reversal of the centrifugal force also at
$r=3M$ (\eg\ \cite{AbramowiczPrasanna1990,1990MNRAS.245..733A},
\cite{Abramowicz1993,1996MNRAS.281..659S}) 
and the fact that compact stars may have moments of inertia
which exceed $\frac{2}{5} M R^2$, the Newtonian value for a sphere
with constant density (\eg\ \cite{1967ApJ...150.1005H,1974MNRAS.167...63C}).

Because the Coriolis force sensitively depends on the ratio of the mass
to the radius of a relativistic star, Type~I X-ray bursts provide a
strong constraint on the equation of state of the underlying neutron
star.  Specifically, since the oscillation frequency of the burst
increases as the atmosphere cools and presumably shrinks, the radius of
the star must exceed $3\; G M/c^2$.  If frame dragging is included, the
constraint is stronger, $R>3.49\; G M/c^2$.
Furthermore, this sensitivity provides a simple explanation for the
fact that the quickly rotating bursters, 4U~1636-54, KS~1731-26, Aql
X-1 and the galactic center source, spin up by the same amount as the
slowly rotating bursters, 4U~1702-43 and 4U~1728-34.  The fast
rotators are simply slightly more massive, 1.6~M$_\odot$ versus
1.4~M$_\odot$.   Further studies of the thermonuclear burning in the
atmospheres of Type~I X-ray bursts will make them a precise probe of
the spacetime geometry surrounding rotating neutron stars.

General relativity presents gravity as an inertial force; therefore,
it is not surprising that Newtonian notions of inertial forces 
do not apply in strong gravitational fields.  Gravity strongly
affects the Coriolis force which is important in the evolution of
Type~I X-Ray bursts; consequently, these bursts provide a strong
constraint on the spacetime geometry surrounding accreting neutron
stars. 

\acknowledgements
I would like to acknowledge a Lee A. DuBridge Postdoctoral Scholarship
for support.  I would like to thank Lars Bildsten, Greg Ushomirsky,
Alessandra Buonanno, Lior Burko, Scott Hughes, Eugene Chiang and Yoram
Lithwick for useful discussions.

\bibliographystyle{jer}
\bibliography{ns,physics,mine,gr}

\end{document}